\documentclass[twocolumn,prb,showpacs,superscriptaddress,showkeys,preprintnumbers,amsmath,amssymb]{revtex4}

\usepackage{graphicx}
\usepackage{dcolumn}
\usepackage{bm}
\usepackage{multirow}
\usepackage{subscript}
\usepackage{nicefrac}

\begin{document}


\title{Incommensurate antiferromagnetic fluctuations in single-crystalline LiFeAs studied by inelastic neutron scattering }

\author{N. Qureshi}
\email[Corresponding author. Electronic
address:~]{qureshi@ph2.uni-koeln.de} \affiliation{$II$.
Physikalisches Institut, Universit\"{a}t zu K\"{o}ln,
Z\"{u}lpicher Strasse 77, D-50937 K\"{o}ln, Germany}

\author{P. Steffens}
\affiliation{Institut Laue Langevin, BP156, 38042 Grenoble Cedex, France}

\author{D. Lamago}
\affiliation{Laboratoire L\'eon Brillouin, CEA/CNRS, F-91191
Gif-sur-Yvette Cedex, France} \affiliation{Institut f\"ur
Festk\"orperphysik, Karlsruher Institut f\"ur Technologie (KIT),
Postfach 3640, D-76121 Karlsruhe, Germany}

\author{Y. Sidis}
\affiliation{Laboratoire L\'eon Brillouin, CEA/CNRS, F-91191 Gif-sur-Yvette Cedex, France}

\author{O. Sobolev}
\affiliation{Institut f\"ur Physikalische Chemie, Georg-August-Universit\"at G\"ottingen, Tammannstrasse 6, 37077 G\"ottingen, Germany }

\author{R. A. Ewings}
\affiliation{ISIS Facility, Rutherford Appleton Laboratory, STFC, Chilton, Didcot, Oxon, OX11 0QX, United Kingdom }

\author{L. Harnagea}
\affiliation{Leibniz-Institute for Solid State Research, IFW-Dresden, 01171 Dresden, Germany}

\author{S. Wurmehl}
\affiliation{Leibniz-Institute for Solid State Research, IFW-Dresden, 01171 Dresden, Germany}
\affiliation{Institut f\"ur Festk\"orperphysik, Technische Universit\"at Dresden, D-01171 Dresden, Germany}

\author{B. B\"uchner}
\affiliation{Leibniz-Institute for Solid State Research, IFW-Dresden, 01171 Dresden, Germany}
\affiliation{Institut f\"ur Festk\"orperphysik, Technische Universit\"at Dresden, D-01171 Dresden, Germany}

\author{M. Braden}
\email{braden@ph2.uni-koeln.de}
\affiliation{$II$.
Physikalisches Institut, Universit\"{a}t zu K\"{o}ln,
Z\"{u}lpicher Strasse 77, D-50937 K\"{o}ln, Germany}

\date{\today}

\begin{abstract}

We present an  inelastic neutron scattering study on
single-crystalline LiFeAs devoted to the characterization of the
incommensurate antiferromagnetic fluctuations at
$\mathbf{Q}=(0.5\pm\delta, 0.5\mp\delta, q_l)$. Time-of-flight
measurements show the presence of these magnetic fluctuations up
to an energy transfer of 60 meV, while polarized neutrons in
combination with longitudinal polarization analysis on a
triple-axis spectrometer prove the pure magnetic origin of this
signal. The normalization of the magnetic scattering to an
absolute scale yields that magnetic fluctuations in LiFeAs are by
a factor eight weaker than the resonance signal in nearly
optimally Co-doped BaFe$_2$As$_2$, although a factor two is
recovered due to the split peaks owing to the incommensurability.
The longitudinal polarization analysis indicates weak spin space
anisotropy with slightly stronger out-of-plane component between 6
and 12 meV. Furthermore, our data suggest a fine structure of the
magnetic signal most likely arising from superposing nesting
vectors.

\end{abstract}


\maketitle

\section{Introduction}
\label{sec:Introduction} The interest in the FeAs-based
superconductors\cite{kam2008} is ongoing after six years of
extensive research as still no consensus has been achieved
concerning the superconducting character and pairing mechanism.
LiFeAs is special amongst the many FeAs-based superconductors, as
superconductivity appears in the parent compound at elevated
temperatures without doping or application of pressure. This
particularity of LiFeAs most likely arises from its electronic
structure with strongly reduced nesting between electron and hole
Fermi-surface sheets as it was first deduced from angle-resolved
photoemission spectroscopy (ARPES) \cite{kor2011}. In the 1111 and
122 families (named after their stoichiometry) the Fermi nesting
conditions are excellent stabilizing a spin density wave (SDW),
which has to be suppressed by
doping\cite{kam2008,lyn2009,rot2008b,sef2008} or the application
of pressure\cite{ali2009} in order to reach the superconducting
state. LiFeAs does not exhibit any structural transition nor a
magnetically ordered phase.\cite{tap2008,wan2008,pit2008}
Theoretical calculations\cite{bry2011} explain this fact by its
poor Fermi nesting properties and unusually shallow hole pockets
around the $\Gamma$ point, which is in agreement with ARPES
experiments.\cite{bor2010,haj2012} The flat top of the hole
pockets implies a large density of states around the $\Gamma$
point and in combination with small-momentum scattering vectors
within the inner hole pocket this would favor ferromagnetic
fluctuations and a triplet pairing mechanism.\cite{bry2011} The
symmetry of the order parameter has been a controversial subject,
several reports using ARPES, quasiparticle interference (QPI) or
theoretical approaches favor an $s^\pm$
wave,\cite{all2012,jan2012,pla2011} while there is also support
for a $p$-wave state.\cite{han2012,bae2012,bae2013} Although the
calculations in Ref.~\onlinecite{pla2011} support an $s^\pm$ wave
state driven by collinear antiferromagnetic fluctuations, the
authors state that ferromagnetic fluctuations stemming from the
small hole pocket at the $\Gamma$ point may dominate at higher
energies and/or at higher temperatures. In our previous
work\cite{qur2012} we have established the energy and temperature
dependence of an antiferromagnetic excitation located at an
incommensurate position $\mathbf{Q}=(0.5\pm\delta, 0.5\mp\delta,
q_l)$ resembling magnetic correlations in electron doped
BaFe$_2$As$_2$. Similar results were obtained by Wang et al.
\cite{wan2012} The origin of the magnetic signal has been
interpreted as scattering between the electron pockets centered
around the $(\pi, \pi)$ point and either the outer\cite{kno2012} or the inner\cite{wan2012} hole pockets around the zone center.\newline
In this work we present a comprehensive inelastic neutron
scattering (INS) study using different cold and thermal
triple-axis spectrometres and a time-of-flight instrument devoted
to extend the characterization of the incommensurate
antiferromagnetic fluctuations in single-crystalline LiFeAs. We
present the inelastic scattered neutron intensity in absolute
units using two different techniques leading to perfectly agreeing
results. The magnetic fluctuations have been investigated up to
energy transfers of 80 meV and spin-space anisotropies have been
studied by polarized neutrons with longitudinal polarization
analysis (LPA). Furthermore, we have investigated
$S(\mathbf{Q},\omega)$ in a broad $\mathbf{Q}$-$\omega$ range to
search for any ferromagnetic fluctuation at elevated temperatures
and energy transfers.

\section{Experimental}
\label{sec:Experimental}

The same single crystal  sample as in Ref.~\onlinecite{qur2012}
has been used for all the experiments presented here. The
normalization to an absolute intensity scale has been done with
data obtained at the thermal triple-axis spectrometer 1T
(Laboratoire L\'eon Brillouin, Saclay), which was used with a
pyrolytic graphite (PG) monochromator and a PG analyzer. The final
neutron energy was fixed at $E_f=14.7$ meV. The IN20 spectrometer
(Institut Laue-Langevin, Grenoble) was used with the FlatCone
multianalyzer in order to record $(hkl)$-maps with different $l$
values at different temperatures and energy transfers. IN20 has
also been used in the polarized mode using polarizing Heusler
(111) crystals as a monochromator and an analyzer. For the LPA a
set of Helmholtz coils was used to guide and orient the neutron
polarization. LPA offers the possibility of distinguishing between
nuclear and magnetic scattering and it furthermore allows the
separation of the two magnetic components perpendicular to the
scattering vector. Generally, nuclear scattering is a
non-spin-flip (NSF) process regardless of the initial neutron
polarization state. Only magnetic components perpendicular to the
scattering vector ($\mathbf{Q}\parallel x$ by definition) are accessible in a neutron experiment. The
components perpendicular to the polarization axis ($y$ being in the scattering plane and $z$ being the perpendicular axis of the spectrometer) contribute to
the spin-flip (SF) channel, while those parallel to the axis of
polarization scatter into the NSF channel. \newline The PUMA
spectrometer (FRM-II, Garching) was used with a PG monochromator
and a PG analyzer with a fixed final neutron energy of $E_f=14.7$
meV. High energy transfers were measured at the time-of flight
spectrometer MAPS (Rutherford-Appleton Laboratory, Didcot). The
incident beam energies were $E_i=55$ and 100 meV with $k_i$
parallel to the $c$ axis. The measured intensities were normalized
to absolute units by using a vanadium standard (with 30\% error).

\section{Results and discussion}
\label{sec:results}

\subsection{Generalized susceptibility of LiFeAs in an absolute scale}
\label{sec:absolute}

In order to express the dynamic susceptibility of LiFeAs in
absolute units data taken on the time-of-flight spectrometer MAPS
and triple-axis spectrometer data from the 1T instrument were used
yielding perfect agreement. The time-of-flight data can be
normalized by comparison with incoherent scattering from a
vanadium sample and with the sample mass. This procedure is
well-established at the MAPS instrument and described in large
detail in reference.\cite{xu2013} In contrast the normalization of
triple-axis data is more complex as the resolution function and
the beam profile are more structured. Here we follow the most
common way to normalize the magnetic scattering by comparison with
phonon measurements on the same sample. This method, furthermore,
excludes mistakes arising from impurity phases.

The scattering potential of the sample is discussed in terms of
the double-differential cross section $ \frac{d^2 \sigma}{d\Omega
dE'}$ with $E'$ the final energy. In any INS experiment this
entity is folded with the resolution  and transmittance function
of the instrument. We use the {\sc reslib} programs\cite{reslib}
to quantitatively analyse the scattering intensities. In our
experiment a neutron monitor between the monochromator and the
sample is used to scale the detector counts into the entity counts
per given monitor (note that this monitor is corrected for higher
order contaminations). The calculation splits the instrumental
effects in the finite Gaussian resolution and a transmittance
term. The intrinsic double differential cross section is first
folded with the Gaussian resolution in the four-dimensional space
consisting of Q-space and energy, see Eq.~\ref{eq:ddcs}. Here
${\bf M}({\bf Q},\omega)$ is the resolution matrix according to
the Popovici approximation \cite{pop1975} and ${\bf \Delta}$ the
four-dimensional difference vector consisting of the Q-space
coordinates and energy, see Eq.~\ref{eq:Delta}.

\begin{align}
\frac{d^2 \tilde{\sigma}}{d\Omega dE'}({\bf Q},\omega)= & \int
\int \int \int  d^3Q' d\omega'  \frac{d^2 \sigma}{d\Omega
dE'}({\bf Q'},\omega') \nonumber\\
& \cdot \exp{\left( -\frac{1}{2}[{\bf \Delta} \cdot{\bf M}({\bf
Q},\omega)\cdot {\bf
\Delta}^T ] \right) }
 \label{eq:ddcs}   \\
 {\bf \Delta} := & (Q'_x-Q_x,Q'_y-Q_y,Q'_z-Q_z,\omega'-\omega)
 \label{eq:Delta}
\end{align}

In order to calculate the intensity in the detector one has to
multiply the folded double cross section with a normalization
factor $R_0({\bf Q},\omega)$ describing amongst others the
efficiency of the secondary spectrometer and the resolution
function normalization $(2\pi)^{-2}\sqrt{\det{\bf M}}$ . In
contrast to the {\sc reslib} manual we do not include the
$\frac{k_f}{k_i}$ factor to $R_0$ but follow the common practice
keeping this factor in the double-differential cross
section.\cite{mar1971,squ1997}

\begin{equation}
I({\bf Q},\omega) = c \cdot R_0({\bf Q},\omega)\cdot \frac{d^2
\tilde{\sigma}}{d\Omega dE'}({\bf Q},\omega) = c \cdot \frac{d^2
\tilde{\tilde{\sigma}}}{d\Omega dE'}({\bf Q},\omega)
 \label{eq:int}
\end{equation}

For known resolution and transmission functions the study of a
predictable signal allows one to determine the scale factor $c$
describing amongst others the effective sample size. The
transformed double-differential cross section $\frac{d^2
\tilde{\tilde{\sigma}}}{d\Omega dE'}({\bf Q},\omega)$ thus
contains the intrinsic scattering strength of the system combined
with the spectrometer properties.

We use the scattering by an acoustic phonon for normalization. The
single-phonon cross section is given by Eq.~\ref{eq:crosssec} (Refs.~\onlinecite{mar1971,squ1997}) where $n(\omega)+1$ is the Bose population factor for neutron energy loss.
$F_{dyn}({\bf Q})$ denotes the dynamical structure factor of the
particular phonon mode at this scattering vector, which can be
calculated with the help of a lattice-dynamical model, see
Eq.~\ref{eq:fdyn}. The $\delta$ function in Eq.~\ref{eq:fdyn2} is
approximated in the calculation by a Lorentzian profile with
finite half width. The symbols in Eqs.~\ref{eq:crosssec}-\ref{eq:fdyn2} follow the same convention as in Refs.~\onlinecite{mar1971,squ1997}.

\begin{align}
\frac{d^2 \sigma}{d\Omega dE'}({\bf Q},\omega) &
=N\cdot\frac{k_f}{k_i}\cdot \frac{n(\omega)+1}{2\omega(q)}\cdot |
F_{dyn}({\bf Q}) |^2 \cdot
\delta(\omega-\omega(q)) \nonumber\\
  = N\cdot & \frac{k_f}{k_i}\cdot
\frac{n(\omega)+1}{2\omega(q)}\cdot | F_{dyn}({\bf Q}) |^2 \cdot
\hbar\delta(E-E(q)) \label{eq:crosssec} \\
F_{dyn}({\bf Q}) & = \sum\limits_{d} \frac{b_d}{\sqrt{m_d}} \cdot
e^{-i{\bf Q}\cdot {\bf r}_d} \cdot {\bf Q}\cdot \hat{\bf e}
(\mathbf{q}) \cdot e^{-W_d(Q)}
\label{eq:fdyn}
\end{align}

For an acoustic phonon close to the Brillouin-zone center one may
further simplify the calculation as all atoms in the primitive
cell are parallel polarized with components
$\frac{\sqrt{m_d}}{\sqrt{M_{tot}}}$ (here $m_d$ and $M_{tot}$
denote the individual and total masses, respectively). The dynamic
structure factor then corresponds to that of the nuclear Bragg
reflection multiplied by the length of the scattering vector, $Q$,
the inverse square root of the total mass and by the cosine of the
angle between scattering vector and phonon polarization,
$cos(\alpha)$. The latter factor is close to one in a reasonably
chosen scan.

\begin{equation}
F_{dyn}({\bf Q}) = \frac{Q\cdot
cos(\alpha)}{\sqrt{M_{tot}}}\sum\limits_{d} b_d \cdot
e^{-i\left[{\bf Q}\cdot {\bf r}_d+W_d(Q)\right]} \label{eq:fdyn2} \end{equation}

\begin{figure}
\includegraphics[width=0.48\textwidth]{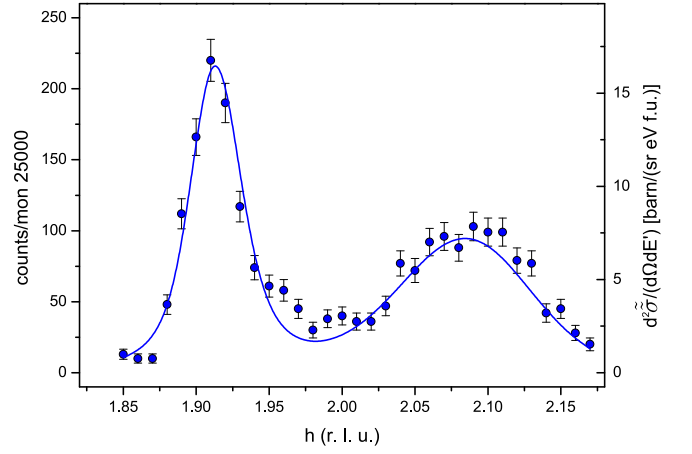}
\caption{\label{fig:phonon} (Color online) Raw data showing the
transversal $\mathbf{q}$-scan across the (220) phonon at $T=3.5$ K
an energy transfer of 4.5 meV measured at the 1T spectrometer. The scattered intensity is given in
counts/(monitor 25000) on the left ordinate and in absolute cross
section values on the right ordinate for the folded cross section
$\frac{d^2 \tilde{\tilde{\sigma}}}{d\Omega dE'}({\bf Q},\omega)$. }
\end{figure}

The double differential cross section of the phonon scattering is
obtained by subtracting the (refined) background from the raw data
and then by dividing by the (refined) scale factor. The phonon
dispersion is described by a simple linear relation, $\omega = c
\cdot |q|$. Fitting the phonon cross-section with its intensity
prefactors to the raw data using the {\sc reslib} code yields a
scale factor of 13.1(8) and a constant background of 2 counts per
monitor. The raw data can therefore be converted into an absolute
scale that still contains the resolution functions of the
instrument, see the right axis of ordinate in
Fig.~\ref{fig:phonon}.

In order to evaluate the magnetic signal we start with the
autocorrelation of the spin Fourier coefficients $S_{\bf
 Q}^\alpha(t)$, here $\alpha, \beta$ denote the space indices,
$\gamma$ the neutron gyromagnetic factor, $e$ the electron charge,
$m_e$ the electron mass, $c$ the speed of light, $f(Q)$ the magnetic form
factor at the scattering vector, $g$ the Land\'e factor and
$\delta_{\alpha \beta}$ the Kronecker symbol. Note that the second
line of Eq.~\ref{eq:ddcsmag} has the unit of an inverse energy
(eV$^{-1}$).

\begin{align}
\frac{d^2 \sigma}{d\Omega dE'}({\bf Q},\omega)  =
\left(\frac{\gamma e^2}{m_e c^2}\right)^2\left(
\frac{gf(Q)}{2}\right)^2\frac{k_f}{k_i}e^{-2W(Q)} \ \ \ \ \ \hfill \nonumber\\
\ \ \  \hfill \cdot \sum\limits_{\alpha \beta}(\delta_{\alpha
 \beta}-\hat{Q}_\alpha \hat{Q}_\beta )\frac{1}{2\pi \hbar}
 \int_{-\infty}^{\infty}dt\cdot e^{-i\omega t}\left<S_{\bf Q}^\alpha(0),S_{\bf
 Q}^\beta(t)\right>
 \label{eq:ddcsmag}
\end{align}

With the fluctuation dissipation theorem  one may transform the
cross section to the imaginary part of the generalized dynamic
susceptibility, which we assume here to be isotropic in spin
space.

\begin{align}
\frac{d^2 \sigma}{d\Omega dE'}&({\bf Q},\omega)   =
\left(\frac{\gamma e^2}{m_e c^2}\right)^2\left(
\frac{gf(Q)}{2}\right)^2\frac{k_f}{k_i}e^{-2W(Q)} \nonumber\\
 & \cdot \frac{N}{\pi(g\mu_B)^2}\left[n(\omega)+1\right]\cdot 2\cdot
 \chi''( {\bf Q},\omega) \nonumber \\
 \label{eq:ddcsmag2}
\end{align}

A susceptibility can be given
in various units creating considerable confusion but here the unit
problem drops out due to the term Bohr-magneton, $\mu_B$, squared
in the denominator. The natural microscopic unit to discus thevsusceptibility is thus $\mu_B^2/eV$ per formula unit, which we
will use in the following.

Deducing the absolute scale of the cross section of the magnetic
fluctuation is now obtained in the same way as in the phonon case
by subtracting the background and by dividing by the scale factor
obtained from the phonon fit. However, in order to fit the data
and deduce the background a model is needed to describe the
generalized susceptibility $\chi''(\mathbf{q},\omega)$. We assume
a superposition of single relaxor functions (Eq.~\ref{eq:relaxor})
in energy with a Lorentzian q-dependence centered at the four
positions ${\bf q_c}$=(0.5$\pm$$\delta_1$,0.5$\mp$$\delta_1$,0)
and (0.5$\pm$$\delta_2$,0.5$\mp$$\delta_2$,0). We take only the
in-plane components of ${\bf q}$ into account. $\delta_1$
($\delta_2$) is the incommensurability of 0.057(3) r.l.u. [0.17(2) r.l.u.] (see
Sec.~\ref{sec:tas} for a detailed description of the two signals)
and the HWHM was refined to $\xi_1$=0.042(9) r.l.u [$\xi_2$=0.07(3)], which
yields the best agreement with the experimental data.

\begin{equation}
\chi''({\bf q},\omega)=\chi'({\bf q_c},0)\cdot\frac{\xi^2}{({\bf
q}-{\bf q_c})^2+\xi^2}\frac{\Gamma \hbar \omega}{(\hbar
\omega)^2+\Gamma^2}
\label{eq:relaxor}
\end{equation}

\begin{figure}
\includegraphics[width=0.48\textwidth]{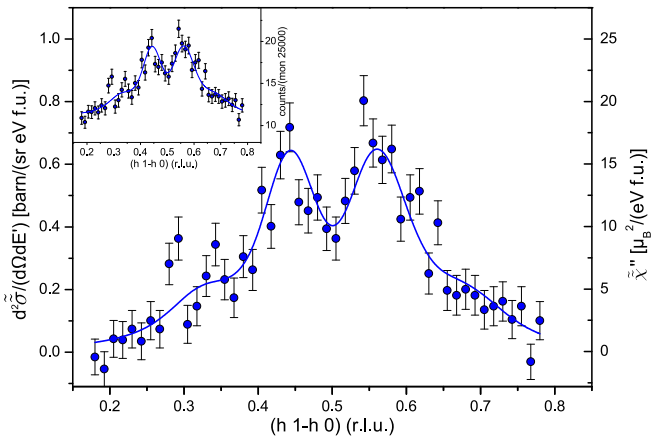}
\caption{\label{fig:absfluct} (Color online) Incommensurate
antiferromagnetic fluctuations at $T=3.5$ K and at an energy
transfer of 5 meV measured at the 1T spectrometer. The ordinates are given in absolute units of
cross section (left) or of the generalized susceptibility (right).
The inset shows the raw magnetic data before subtraction of the
background and the division by the scale factor.}
\end{figure}

In a first step, the constant background (11 counts per monitor
25000 $\sim$ 20 s) was determined and subtracted from the raw data
which was then divided by the scale factor deduced from the phonon
fit yielding the transformed double-differential cross section
$\frac{d^2 \tilde{\tilde{\sigma}}}{d\Omega dE'}({\bf Q},\omega)$.
From this one may obtain a susceptibility folded with the
instrument resolution and transmission by dividing by all the
intensity prefactors in Eq.~\ref{eq:ddcsmag2}; this result is
shown in Fig. 2 on the right coordinate axis. The intrinsic
strength and shape of $\chi''(\mathbf{q},\omega)$, however, can
only be obtained by fitting the model, Eq.~\ref{eq:relaxor}, to
the raw data (a value of 10.9 meV has been used for $\Gamma$
obtained by the single-relaxor fit to the data presented in
Sec.~\ref{sec:results}.B). Thereby we obtain $\chi'({\bf
q_c},0)$=7.4(8) and \mbox{35(1) $\mu_B^2/eV$} at the outer and inner
incommensurate positions, respectively, see Fig.~\ref{fig:absfluct}.

\subsection{Evidence for additional contributions to magnetic scattering}
\label{sec:tas}

\begin{figure*}
\includegraphics[width=0.7\textwidth]{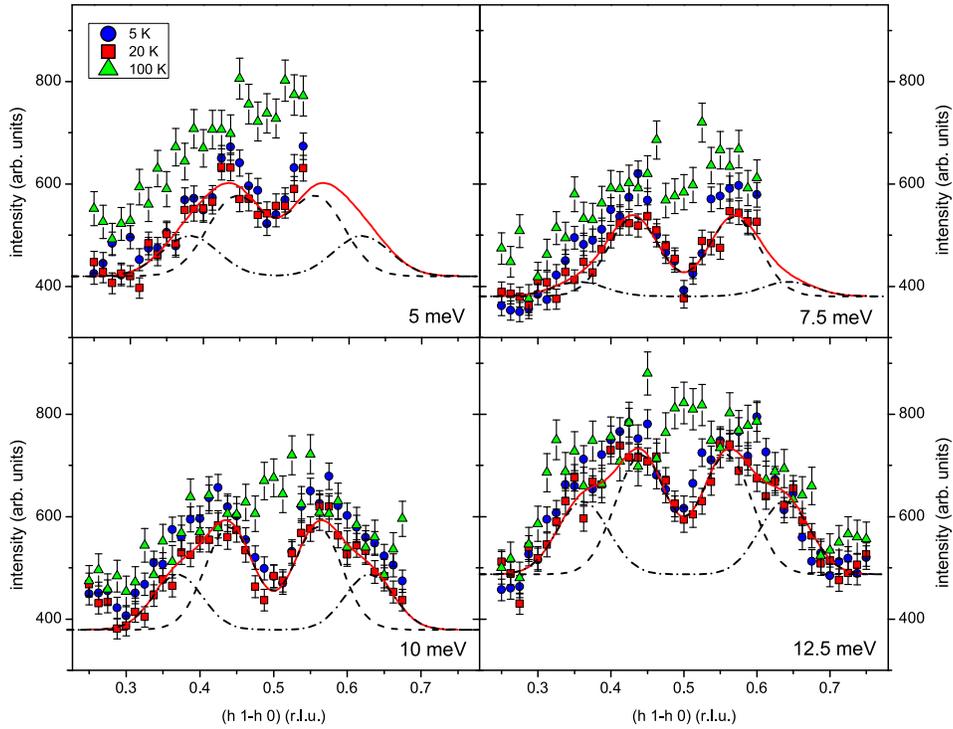}
\caption{\label{fig:puma} (Color online) Transverse q-scans across (0.5 0.5 0) at different energy transfers measured at the PUMA spectrometer. The data have been fitted by two pairs of symmetrical Gaussians on a constant background. A phonon contamination is visible towards smaller energy transfer, for which those data points have been omitted. Especially at higher energy transfers the data indicate an additional signal at larger incommensurability (dash-dotted line) besides the signal at (0.43 0.57 0) (dashed line). }
\end{figure*}

The incommensurate fluctuation has been reinvestigated at the
thermal triple-axis spectrometer PUMA (FRM-II, Garching).
Fig.~\ref{fig:puma} shows transverse q-scans across the (0.5 0.5
0) position at different energy transfer clearly documenting a
complex Q-shape of the magnetic response. The incommensurate
magnetic correlations exhibit at least an asymmetric profile with
pronounced shoulders towards larger incommensurability (compared
to the (0.5,0.5,$q_l$) center). Therefore, the data have been
described with two pairs of symmetrical Gaussian functions on a
constant background. Note that parts of the data are contaminated
by phonon scattering towards lower energy transfer for which these
data points are not shown. The resulting fit curves [(red) solid
lines]) show a very good agreement with the raw data, while the
dashed (dash-dotted) curves indicate the contribution of the
signal at $Q_{inc,1}$$\approx$(0.43 0.57 0)
[$Q_{inc,2}$$\approx$(0.35 0.65 0)].

Angle resolved photoemission spectroscopy (ARPES) experiments
\cite{bor2010,kor2011} have revealed the Fermi surface to consist
of two similarly sized electron-like sheets around the $X$ point
and  hole-like sheets around the $\Gamma$ point. In
Ref.~\onlinecite{kno2012} the authors have identified the INS
signal to be connected to scattering between the outer hole pocket
and the electron pockets by using a simple tight-binding fit to
the ARPES data, while involvement of the inner hole pocket was concluded in Ref.~\onlinecite{wan2012}. Deeper understanding of the nesting signal
requires the analysis of the orbital character of the various
Fermi surface sheets which essentially arise from the $t_{2g}$
$d_{xz}$, $d_{yz}$ and $d_{xy}$ orbitals.\cite{yin2011,fer2012} There
seems to be agreement that the outer hole pocket can be identified
with $d_{xy}$ orbital character which also contributes to the
electron pockets. The $d_{xy}$ states should result in
two-dimensional bands, but $d_{xz}$ and $d_{yz}$ contributions
yield considerable dispersion along the perpendicular directions
and strong $q_z$ modulation of the Fermi surfaces.\cite{yin2011,fer2012}
If one associates the nesting magnetic correlations exclusively
with $d_{xy}$ orbitals it appears difficult to understand a split
signal but an asymmetry or a shoulder can arise from a peculiar
detail of the Fermi surface shape that is not sufficiently well
understood so far. Sr$_2$RuO$_4$ is a well studied example with
incommensurate magnetic correlations arising from Fermi-surface
nesting,\cite{bra2002} and this material also exhibits an
asymmetric magnetic response with a shoulder. Quite recently four
theoretical papers aimed to quantitatively model the variation of
the superconducting gap on the Fermi surface sheets arriving at
contradictory results.\cite{wan2013,yin2013,ahn2014,sai2014} The
quantitative description of magnetic excitations by analyzing
transitions between states with the same or different orbital
character will help to arrive at a better understanding of the
electronic structure of LiFeAs.

\begin{figure}
\includegraphics[width=0.42\textwidth]{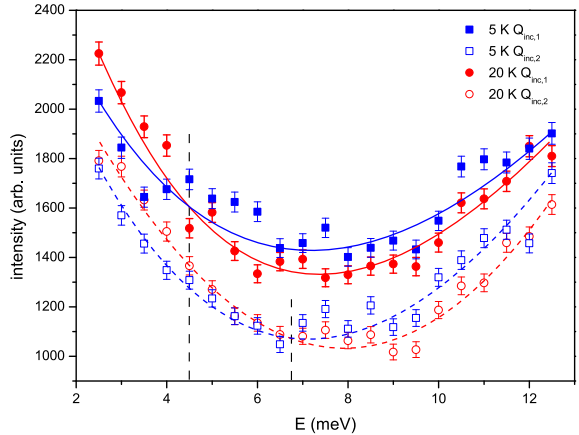}
\caption{\label{fig:escanpuma} (Color online) Energy dependence of the INS scattering at $Q_{inc,1}$=(0.425 0.575 0) and $Q_{inc,2}$=(0.35 0.65 0) measured at the PUMA spectrometer. The dashed lines indicate where the scattered intensities of the normal [(red) circles)] and superconducting state [(blue) squares] cross.}
\end{figure}

The peak intensities of $Q_{inc,1}$ and $Q_{inc,2}$ have been
followed as a function of energy above and below $T_C$. As it can
be seen in Fig.~\ref{fig:escanpuma} the main signal $Q_{inc,1}$
shows the same dependence as already reported in
Ref.~\onlinecite{qur2012} with a crossover between the scattered
intensity below and above $T_C$ at 4.5 meV and an increase of
intensity above 7-8 meV. On the other hand the intensity at
$Q_{inc,2}$ suggests a different behaviour in dependence on the
energy transfer. The scattered intensity at 20 K stays above the
one at 5 K up to an energy transfer of roughly 7 meV, above which the value
$I(T$$<$$T_C)$ becomes stronger than $I(T$$>$$T_C)$. The different
energy dependences of the signals at $Q_{inc,1}$ and $Q_{inc,2}$
strengthen the assumption of their independent origin and can be
explained due to different gap values on different parts of the
Fermi surface.

\begin{figure}
\includegraphics[width=0.42\textwidth]{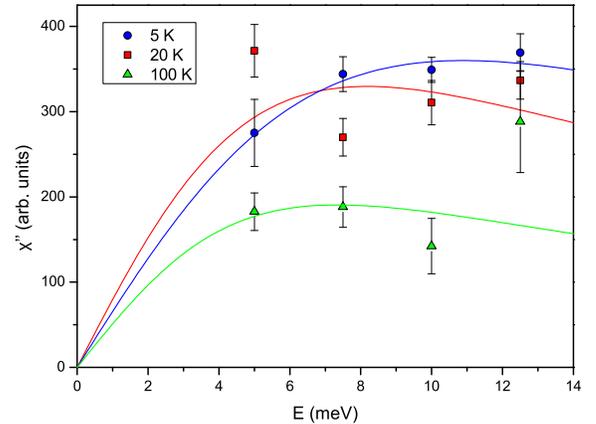}
\caption{\label{fig:relaxor} (Color online) Imaginary part of the generalized susceptibility at 5 K, 20 K and 100 K as obtained by the amplitude from the fits to the data shown in Fig.~\ref{fig:puma} and correction for the monitor and the Bose factor. The solid lines represent fits by a single relaxor functions $\chi''(Q_{inc,1},E)=\chi'(Q_{inc,1},0)\frac{\Gamma E}{\Gamma^2+E^2}$}
\end{figure}

The derived amplitude of the excitation at $Q_{inc,1}$ has been
corrected for the monitor and the Bose factor yielding the
imaginary part of the generalized susceptibility which is shown in
Fig.~\ref{fig:relaxor}. The data have been fitted with
single-relaxor functions. The data does not allow to state a clear
tendency of the critical energy, however, a clear reduction of
$\chi''(Q_{inc},E)$ towards higher temperatures is observable. In
addition the incommensurate magnetic correlations become strongly
broadened at the temperature of only 100\ K where the two peak
structure has already changed into a broad plateau.

By using the triple-axis spectrometer IN20 in combination with the Flatcone multianalyzer, planar sections of the reciprocal space can be recorded by simple $2\theta$ scans which are afterwards converted into $\mathbf{Q}$-space.
As theory predicts that a ferromagnetic instability may dominate at higher temperatures and/or higher energies,\cite{pla2011} maps of the reciprocal space have been recorded up to 150 K and an energy transfer of 40 meV focusing on the (100) and (110) positions. However, our obtained data does not give any hint for ferromagnetic fluctuations in LiFeAs.

\subsection{Spin space anisotropy of magnetic correlations}
\label{sec:spinspace}

\begin{figure}
\includegraphics[width=0.42\textwidth]{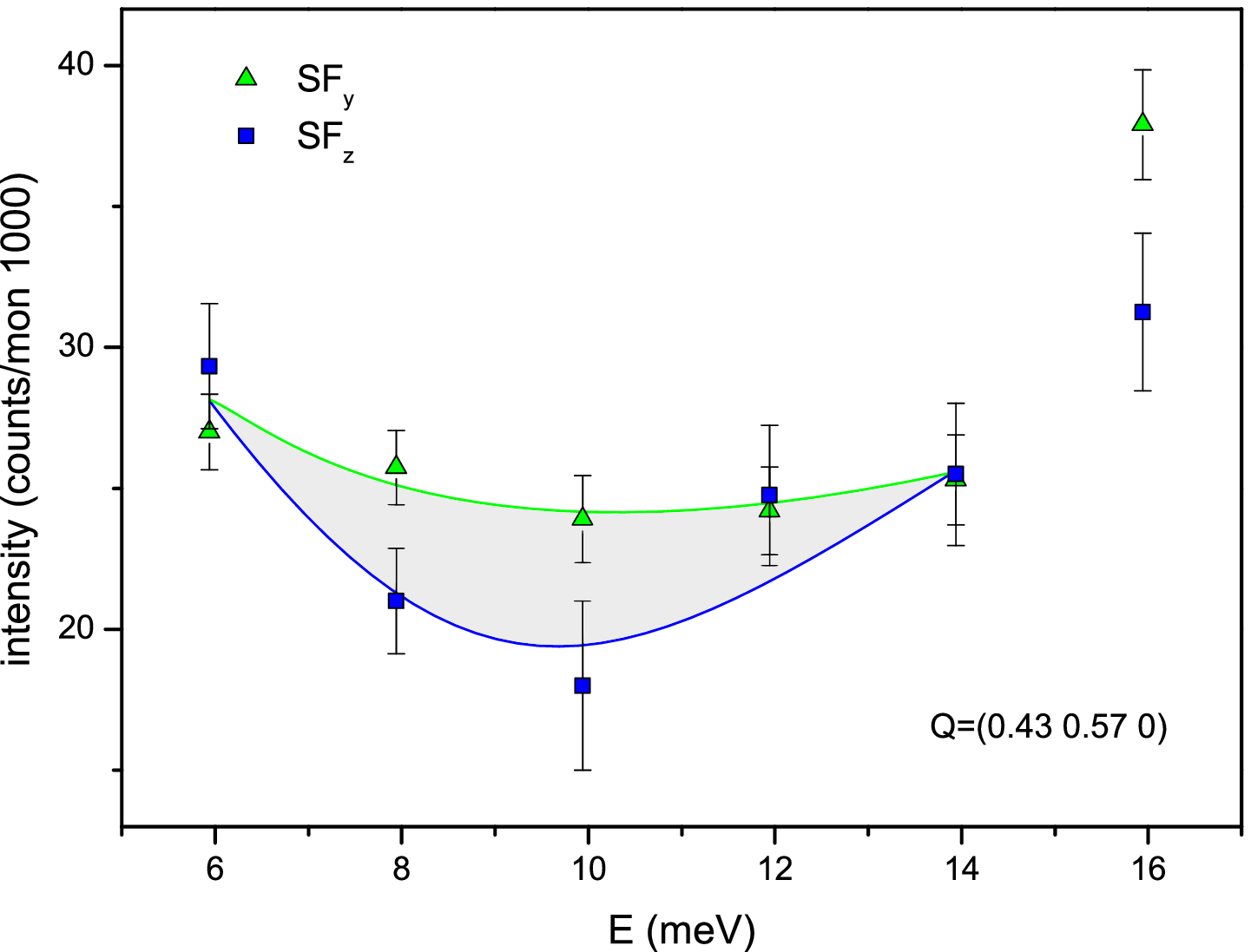}
\caption{\label{fig:escan} (Color online) Energy scan of the SF$_y$ and SF$_z$ intensities at $\mathbf{Q}$=(0.43 0.57 0) measured at the IN20 spectrometer showing a local anisotropy of the magnetic fluctuation between 6 and 12 meV.}
\end{figure}

The IN20 spectrometer has then been used with polarized neutrons
whose polarization axis after the scattering process has been
analyzed. The observation of the incommensurate signal in the SF
channels proves its magnetic origin (Note that the SF background
has been subtracted according to the description in
Ref.~\onlinecite{qur2012b}).
The peak intensity at the point $\mathbf{Q}$=(0.43 0.57 0) has
been measured as a function of the energy transfer for the SF$_y$
and SF$_z$ channels (Fig.~\ref{fig:escan}). Although only the
SF$_y$ channel has been measured with high statistics a slight
spin-space anisotropy of the magnetic fluctuation is visible
between 6 and 12 meV, where the out-of-plane fluctuation lies
above the in-plane fluctuation similar to observations in electron
doped BaFe$_2$As$_2$ \cite{ste2013,liu2012,luo2013}. However, the spin-space
anisotropy in LiFeAs needs further experimental corroboration by
measuring the other channels with better statistics.

In order to reveal eventual weak ferromagnetic fluctuations a
$\mathbf{q}$-scan across $\mathbf{Q}$=(110) at $T=150$ K and
$E=12$ meV has been carried out. All three SF channels revealed
neutron counts similar to the SF background meaning that no
significant magnetic scattering is present.

\subsection{High-energy magnetic response}
\label{sec:unpolneutron}

\begin{figure}
\includegraphics[width=0.42\textwidth]{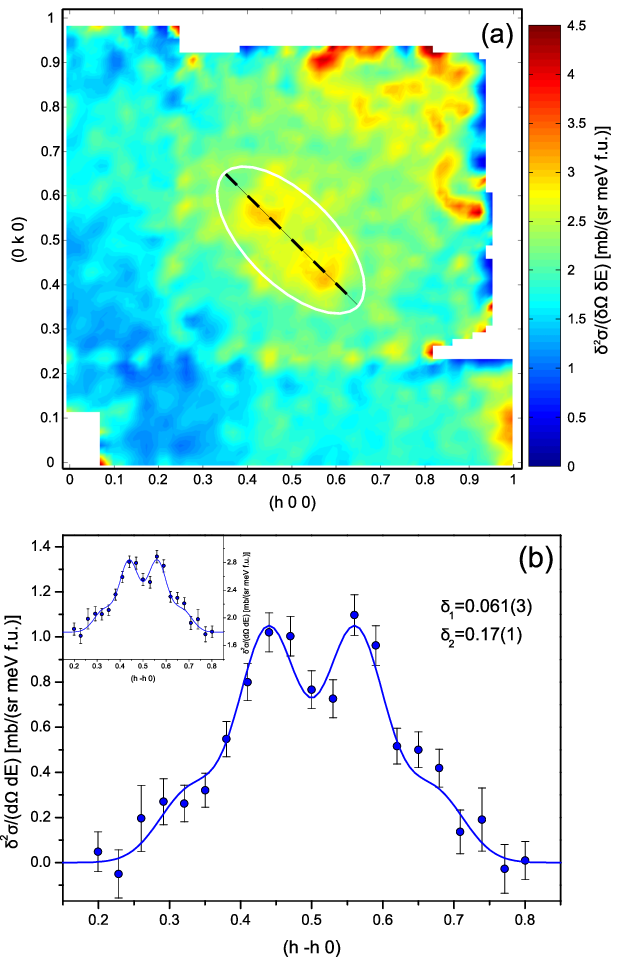}
\caption{\label{fig:fluctuation} (Color online) (a) Time-of-flight
data showing $S(\mathbf{Q},\omega)$ in the $(hk0)$ plane
integrated between 10 meV and 30 meV ($E_i$=55 meV, $T$=10 K. The
incommensurate peaks are marked by a white ellipse. The black
dashed indicates the cut along the [h -h 0] direction shown in
(b). The fit of two pairs of symmetrical Gaussian functions after
the subtraction of a constant background (determined from the fit
in the inset) yields the incommensurabilities $\delta_1=0.061(3)$
and $\delta_2=0.17(1)$.}
\end{figure}

\begin{figure*}
\includegraphics[width=0.9\textwidth]{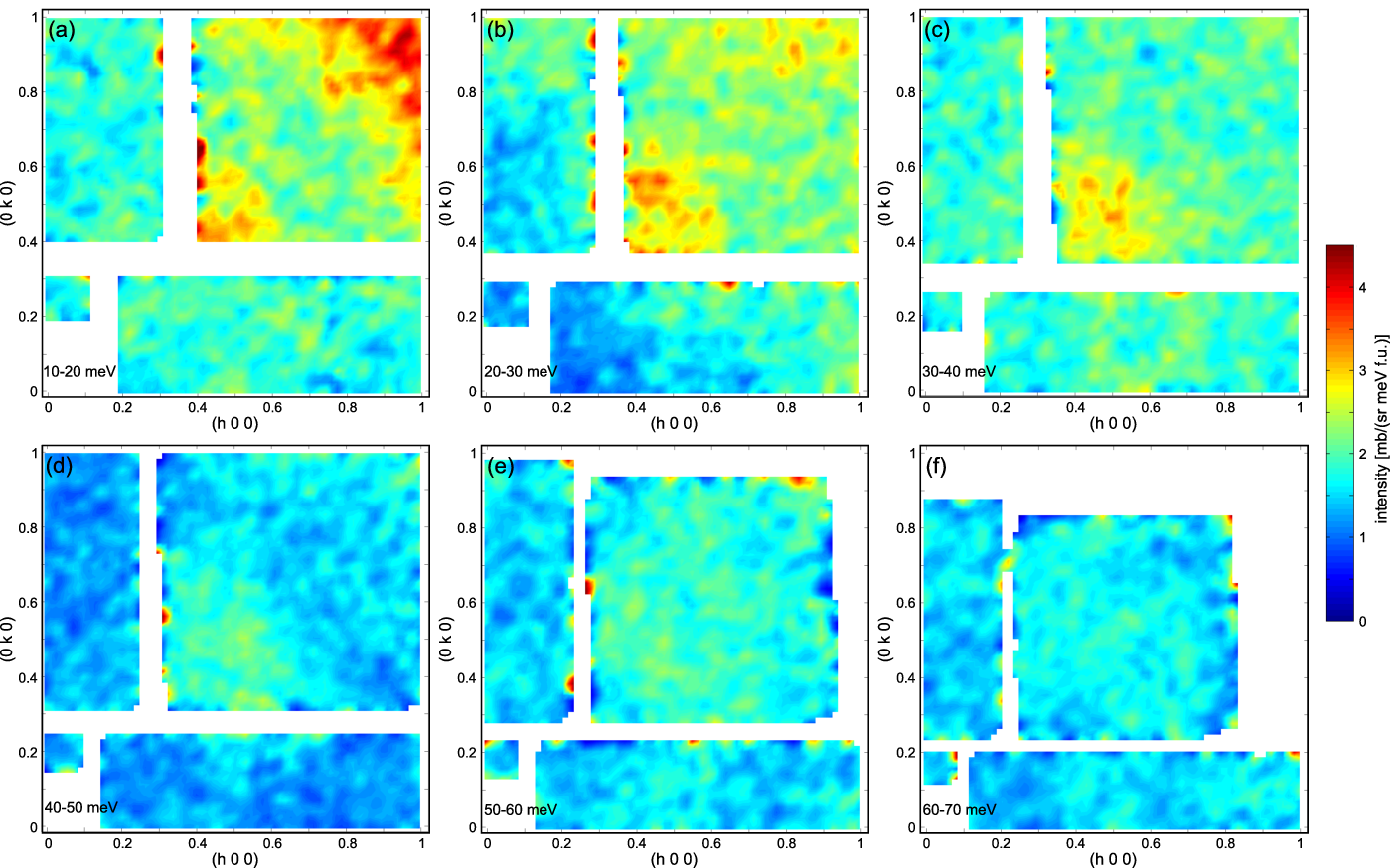}
\caption{\label{fig:E100} (Color online) (a)-(f)
$S(\mathbf{Q},\omega)$ in different energy slices from 10-20 meV
up to 60-70 meV ($E_i$=100 meV, $T$=10 K). The magnetic signal is
observable up to 50-60 meV before it is reduced to the background
at higher energies.}
\end{figure*}

Due to the limitation of triple-axis spectrometers concerning the
incident energy, high energy transfers have to be measured using a
time-of-flight spectrometer. However, higher incident energies are
at the cost of a loss in resolution. 
With the $c$ axis of the sample aligned along the incident beam one obtains a projection of $S(\mathbf{Q},\omega)$ along this axis after the measurement of a curved 3-dimensional hypersurface in the 4-dimensional manifold of reciprocal space. In the projection the $l$ component is an implicit variable which changes with energy tranfer, nevertheless being calculable, i.e. the obtained data is
three-dimensional in $(h,k,E)$-space. In order to visualize the
data the program {\sc mslice} has been used which offers the
possibility of averaging the data along a chosen axis to produce
a slice or integrating along two axes to produce a cut. By
measuring a standard vanadium sample with known mass the intensity
can be normalized to an absolute scale in mb/(sr meV f.u.) by
using the sample mass and molar mass.
Fig.~\ref{fig:fluctuation}(a) shows a slice of the $(hk0)$ plane
which has been integrated between 10 meV and 30 meV for an
incident beam energy of 55 meV. Two peaks can be observed around
the (0.5 0.5 0) position. By integrating the data perpendicular to
the scan path indicated by the dashed line, one obtains the curve
shown in Fig.~\ref{fig:fluctuation}(b) clearly revealing the
incommensurability of the antiferromagnetic fluctuations. However,
also the time-of-flight data indicates an asymmetric shape or an
additional signal at larger incommensurability. Two pairs of
symmetrical Gaussian functions on a constant background have been
fitted to the data, from which the incommensurabilities
$\delta_1=0.061(3)$ and $\delta_2=0.17(1)$ could be extracted. The
value of $\delta_1$ is in good agreement with our previous
results.\cite{qur2012} Note that the absolute intensity scale
obtained by the renormalization to a vanadium standard (between 10
and 30 meV) is in very good agreement with our results shown in
Sec.~\ref{sec:absolute} (Fig.~\ref{fig:absfluct}, 5 meV) and also with a report on polycrystalline samples.\cite{tay2011}
In order to investigate the magnetic signal at higher energy
transfers an incident neutron energy of 100 meV has been used.
Fig.~\ref{fig:E100} shows $(hk0)$ slices of 10 meV thickness each.
The magnetic fluctuation can be observed around the (0.5 0.5 0)
point, but the loss in resolution becomes evident. However, there
is a significant signal which can be separated from the background
up to an energy transfer of 60 meV. For the slice in
Fig.~\ref{fig:E100}(f) the signal is reduced to the background.
Due to the limited $\mathbf{Q}$-resolution in comparison to
$E_i$=55 meV the incommensurability could not be investigated at
higher energy transfers.

\section{Conclusion}
\label{sec:discussion}

In summary we have extended our previous work concerning the
characterization of the incommensurate antiferromagnetic
fluctuations in LiFeAs. Time-of-flight experiments show that the
magnetic signal is observable up to energy transfers of 60 meV,
while the incommensurability remains unchanged up to 30 meV
(measurements of higher energy transfers were at the cost of
resolution prohibiting a quantitative analysis of the
incommensurability). Longitudinal polarization analysis proved the
magnetic origin of the observed signal and an eventual spin-space
anisotropy between 8 and 10 meV could be deduced that resembles
observation in other FeAs-based superconductors.

The asymmetric shape of the incommensurate peak suggests the
presence of two different signals which may correspond to
scattering between the outer hole pocket and the inner electron
pocket as well as between the outer hole pocket and the outer
electron pocket. The different energy dependences of the peak
intensities of $Q_{inc,1}$ and $Q_{inc,2}$ support the picture of
two independent signals. Furthermore, we have converted the
intensity of the scattered neutrons into an absolute scale making
it possible to compare the strength of the magnetic fluctuations
in LiFeAs with those of related compounds. Nearly optimally
Co-doped BaFe$_2$As$_2$ yields a maximum value of roughly 7.5
mb/(sr meV Fe) at the resonance feature
(Ref.~\onlinecite{les2010}). By averaging the peak values of the
triple-axis (Fig.~\ref{fig:absfluct}) and time-of-flight data
(Fig.~\ref{fig:fluctuation}) we obtain 0.95 mb/(sr meV Fe)
rendering the low-temperature fluctuations in LiFeAs by a factor 8
weaker than the magnetic resonance in  Co-doped BaFe$_2$As$_2$.
This perfectly agrees with our earlier work,\cite{qur2012} where
we estimated the same ratio between the incommensurate
fluctuations in LiFeAs and the commensurate resonance in
Ba(Fe$_{0.92}$Co$_{0.08}$)$_2$As$_2$ by normalizing the magnetic
signal to the respective phonon signal. Due to the
incommensurability a factor of two is recovered for which the
magnetic scattering per Fe ion in LiFeAs is by roughly a factor
four weaker than the respective scattering in nearly optimally
Co-doped BaFe$_2$As$_2$ which must be reconciled with the fact
that the superconducting transition temperature is only little
reduced in LiFeAs.

\begin{acknowledgments}
This work was supported by the Deutsche Forschungsgemeinschaft
(DFG) through the Priority Programme SPP1458 (Grants No.
BE1749/13, BU887/15-1 and BR2211/1-1). S.W. thanks the DFG for
funding in the Emmy Noether Programme (project 595/3-1).

\end{acknowledgments}



\begin{thebibliography}{39}
\expandafter\ifx\csname natexlab\endcsname\relax\def\natexlab#1{#1}\fi
\expandafter\ifx\csname bibnamefont\endcsname\relax
  \def\bibnamefont#1{#1}\fi
\expandafter\ifx\csname bibfnamefont\endcsname\relax
  \def\bibfnamefont#1{#1}\fi
\expandafter\ifx\csname citenamefont\endcsname\relax
  \def\citenamefont#1{#1}\fi
\expandafter\ifx\csname url\endcsname\relax
  \def\url#1{\texttt{#1}}\fi
\expandafter\ifx\csname urlprefix\endcsname\relax\def\urlprefix{URL }\fi
\providecommand{\bibinfo}[2]{#2}
\providecommand{\eprint}[2][]{\url{#2}}

\bibitem[{\citenamefont{Kamihara et~al.}(2008)\citenamefont{Kamihara, Watanabe,
  Hirano, and Hosono}}]{kam2008}
\bibinfo{author}{\bibfnamefont{Y.}~\bibnamefont{Kamihara}},
  \bibinfo{author}{\bibfnamefont{T.}~\bibnamefont{Watanabe}},
  \bibinfo{author}{\bibfnamefont{M.}~\bibnamefont{Hirano}}, \bibnamefont{and}
  \bibinfo{author}{\bibfnamefont{H.}~\bibnamefont{Hosono}},
  \bibinfo{journal}{J. Am. Chem. Soc.} \textbf{\bibinfo{volume}{130}},
  \bibinfo{pages}{3296} (\bibinfo{year}{2008}).

\bibitem[{\citenamefont{Kordyuk et~al.}(2011)\citenamefont{Kordyuk, Zabolotnyy,
  Evtushinsky, Kim, Morozov, Kulic, Follath, Behr, B\"uchner, and
  Borisenko}}]{kor2011}
\bibinfo{author}{\bibfnamefont{A.~A.} \bibnamefont{Kordyuk}},
  \bibinfo{author}{\bibfnamefont{V.~B.} \bibnamefont{Zabolotnyy}},
  \bibinfo{author}{\bibfnamefont{D.~V.} \bibnamefont{Evtushinsky}},
  \bibinfo{author}{\bibfnamefont{T.~K.} \bibnamefont{Kim}},
  \bibinfo{author}{\bibfnamefont{I.~V.} \bibnamefont{Morozov}},
  \bibinfo{author}{\bibfnamefont{M.~L.} \bibnamefont{Kulic}},
  \bibinfo{author}{\bibfnamefont{R.}~\bibnamefont{Follath}},
  \bibinfo{author}{\bibfnamefont{G.}~\bibnamefont{Behr}},
  \bibinfo{author}{\bibfnamefont{B.}~\bibnamefont{B\"uchner}},
  \bibnamefont{and} \bibinfo{author}{\bibfnamefont{S.~V.}
  \bibnamefont{Borisenko}}, \bibinfo{journal}{Phys. Rev. B}
  \textbf{\bibinfo{volume}{83}}, \bibinfo{pages}{134513}
  (\bibinfo{year}{2011}).

\bibitem[{\citenamefont{Lynn. and Dai}(2009)}]{lyn2009}
\bibinfo{author}{\bibfnamefont{J.~W.} \bibnamefont{Lynn.}} \bibnamefont{and}
  \bibinfo{author}{\bibfnamefont{P.~C.} \bibnamefont{Dai}},
  \bibinfo{journal}{Physica (Amsterdam)} \textbf{\bibinfo{volume}{469C}},
  \bibinfo{pages}{469} (\bibinfo{year}{2009}).

\bibitem[{\citenamefont{Rotter et~al.}(2008)\citenamefont{Rotter, Tegel, and
  Johrendt}}]{rot2008b}
\bibinfo{author}{\bibfnamefont{M.}~\bibnamefont{Rotter}},
  \bibinfo{author}{\bibfnamefont{M.}~\bibnamefont{Tegel}}, \bibnamefont{and}
  \bibinfo{author}{\bibfnamefont{D.}~\bibnamefont{Johrendt}},
  \bibinfo{journal}{Phys. Rev. Lett.} \textbf{\bibinfo{volume}{101}},
  \bibinfo{pages}{107006} (\bibinfo{year}{2008}).

\bibitem[{\citenamefont{Sefat et~al.}(2008)\citenamefont{Sefat, Jin, McGuire,
  Sales, Singh, and Mandrus}}]{sef2008}
\bibinfo{author}{\bibfnamefont{A.~S.} \bibnamefont{Sefat}},
  \bibinfo{author}{\bibfnamefont{R.}~\bibnamefont{Jin}},
  \bibinfo{author}{\bibfnamefont{M.~A.} \bibnamefont{McGuire}},
  \bibinfo{author}{\bibfnamefont{B.~C.} \bibnamefont{Sales}},
  \bibinfo{author}{\bibfnamefont{D.~J.} \bibnamefont{Singh}}, \bibnamefont{and}
  \bibinfo{author}{\bibfnamefont{D.}~\bibnamefont{Mandrus}},
  \bibinfo{journal}{Phys. Rev. Lett.} \textbf{\bibinfo{volume}{101}},
  \bibinfo{pages}{117004} (\bibinfo{year}{2008}).

\bibitem[{\citenamefont{Alireza et~al.}(2009)\citenamefont{Alireza, Ko,
  J.~Gillett, Cole, Lonzarich, and Sebastian}}]{ali2009}
\bibinfo{author}{\bibfnamefont{P.}~\bibnamefont{Alireza}},
  \bibinfo{author}{\bibfnamefont{Y.~T.~C.} \bibnamefont{Ko}},
  \bibinfo{author}{\bibfnamefont{C.~M.~P.} \bibnamefont{J.~Gillett}},
  \bibinfo{author}{\bibfnamefont{J.~M.} \bibnamefont{Cole}},
  \bibinfo{author}{\bibfnamefont{G.~G.} \bibnamefont{Lonzarich}},
  \bibnamefont{and} \bibinfo{author}{\bibfnamefont{S.~E.}
  \bibnamefont{Sebastian}}, \bibinfo{journal}{J. Phys.: Condens. Matter}
  \textbf{\bibinfo{volume}{21}}, \bibinfo{pages}{012208}
  (\bibinfo{year}{2009}).

\bibitem[{\citenamefont{Tapp et~al.}(2008)\citenamefont{Tapp, Tang, Lv, Sasmal,
  Lorenz, Chu, and Guloy}}]{tap2008}
\bibinfo{author}{\bibfnamefont{J.~H.} \bibnamefont{Tapp}},
  \bibinfo{author}{\bibfnamefont{Z.}~\bibnamefont{Tang}},
  \bibinfo{author}{\bibfnamefont{B.}~\bibnamefont{Lv}},
  \bibinfo{author}{\bibfnamefont{K.}~\bibnamefont{Sasmal}},
  \bibinfo{author}{\bibfnamefont{B.}~\bibnamefont{Lorenz}},
  \bibinfo{author}{\bibfnamefont{P.~C.~W.} \bibnamefont{Chu}},
  \bibnamefont{and} \bibinfo{author}{\bibfnamefont{A.~M.} \bibnamefont{Guloy}},
  \bibinfo{journal}{Phys. Rev. B} \textbf{\bibinfo{volume}{78}},
  \bibinfo{pages}{060505(R)} (\bibinfo{year}{2008}).

\bibitem[{\citenamefont{Wang et~al.}(2008)\citenamefont{Wang, Liu, Lv, Gao,
  Yang, Yu, Li, and Jin}}]{wan2008}
\bibinfo{author}{\bibfnamefont{X.~C.} \bibnamefont{Wang}},
  \bibinfo{author}{\bibfnamefont{Q.~Q.} \bibnamefont{Liu}},
  \bibinfo{author}{\bibfnamefont{Y.}~\bibnamefont{Lv}},
  \bibinfo{author}{\bibfnamefont{W.~B.} \bibnamefont{Gao}},
  \bibinfo{author}{\bibfnamefont{L.}~\bibnamefont{Yang}},
  \bibinfo{author}{\bibfnamefont{R.~C.} \bibnamefont{Yu}},
  \bibinfo{author}{\bibfnamefont{F.~Y.} \bibnamefont{Li}}, \bibnamefont{and}
  \bibinfo{author}{\bibfnamefont{C.~Q.} \bibnamefont{Jin}},
  \bibinfo{journal}{Solid State Commun.} \textbf{\bibinfo{volume}{148}},
  \bibinfo{pages}{538} (\bibinfo{year}{2008}).

\bibitem[{\citenamefont{Pitcher et~al.}(2008)\citenamefont{Pitcher, Parker,
  Adamson, Herkelrath, Boothroyd, Ibberson, Brunelli, and Clarke}}]{pit2008}
\bibinfo{author}{\bibfnamefont{M.~J.} \bibnamefont{Pitcher}},
  \bibinfo{author}{\bibfnamefont{D.~R.} \bibnamefont{Parker}},
  \bibinfo{author}{\bibfnamefont{P.}~\bibnamefont{Adamson}},
  \bibinfo{author}{\bibfnamefont{S.~J.~C.} \bibnamefont{Herkelrath}},
  \bibinfo{author}{\bibfnamefont{A.~T.} \bibnamefont{Boothroyd}},
  \bibinfo{author}{\bibfnamefont{R.~M.} \bibnamefont{Ibberson}},
  \bibinfo{author}{\bibfnamefont{M.}~\bibnamefont{Brunelli}}, \bibnamefont{and}
  \bibinfo{author}{\bibfnamefont{S.~J.} \bibnamefont{Clarke}},
  \bibinfo{journal}{Chem. Commun.} p. \bibinfo{pages}{5918}
  (\bibinfo{year}{2008}).

\bibitem[{\citenamefont{Brydon et~al.}(2011)\citenamefont{Brydon, Daghofer,
  Timm, and van~den Brink}}]{bry2011}
\bibinfo{author}{\bibfnamefont{P.~M.~R.} \bibnamefont{Brydon}},
  \bibinfo{author}{\bibfnamefont{M.}~\bibnamefont{Daghofer}},
  \bibinfo{author}{\bibfnamefont{C.}~\bibnamefont{Timm}}, \bibnamefont{and}
  \bibinfo{author}{\bibfnamefont{J.}~\bibnamefont{van~den Brink}},
  \bibinfo{journal}{Phys. Rev. B} \textbf{\bibinfo{volume}{83}},
  \bibinfo{pages}{060501(R)} (\bibinfo{year}{2011}).

\bibitem[{\citenamefont{Borisenko et~al.}(2010)\citenamefont{Borisenko,
  Zabolotnyy, Evtushinsky, Kim, Morozov, Yaresko, Kordyuk, Behr, Vasiliev,
  Follath et~al.}}]{bor2010}
\bibinfo{author}{\bibfnamefont{S.~V.} \bibnamefont{Borisenko}},
  \bibinfo{author}{\bibfnamefont{V.~B.} \bibnamefont{Zabolotnyy}},
  \bibinfo{author}{\bibfnamefont{D.~V.} \bibnamefont{Evtushinsky}},
  \bibinfo{author}{\bibfnamefont{T.~K.} \bibnamefont{Kim}},
  \bibinfo{author}{\bibfnamefont{I.~V.} \bibnamefont{Morozov}},
  \bibinfo{author}{\bibfnamefont{A.~N.} \bibnamefont{Yaresko}},
  \bibinfo{author}{\bibfnamefont{A.~A.} \bibnamefont{Kordyuk}},
  \bibinfo{author}{\bibfnamefont{G.}~\bibnamefont{Behr}},
  \bibinfo{author}{\bibfnamefont{A.}~\bibnamefont{Vasiliev}},
  \bibinfo{author}{\bibfnamefont{R.}~\bibnamefont{Follath}},
  \bibnamefont{et~al.}, \bibinfo{journal}{Phys. Rev. Lett}
  \textbf{\bibinfo{volume}{105}}, \bibinfo{pages}{067002}
  (\bibinfo{year}{2010}).

\bibitem[{\citenamefont{Hajiri et~al.}(2012)\citenamefont{Hajiri, Ito, Niwa,
  Matsunami, Min, Kwon, and Kimura}}]{haj2012}
\bibinfo{author}{\bibfnamefont{T.}~\bibnamefont{Hajiri}},
  \bibinfo{author}{\bibfnamefont{T.}~\bibnamefont{Ito}},
  \bibinfo{author}{\bibfnamefont{R.}~\bibnamefont{Niwa}},
  \bibinfo{author}{\bibfnamefont{M.}~\bibnamefont{Matsunami}},
  \bibinfo{author}{\bibfnamefont{B.~H.} \bibnamefont{Min}},
  \bibinfo{author}{\bibfnamefont{Y.~S.} \bibnamefont{Kwon}}, \bibnamefont{and}
  \bibinfo{author}{\bibfnamefont{S.}~\bibnamefont{Kimura}},
  \bibinfo{journal}{Phys. Rev. B} \textbf{\bibinfo{volume}{85}},
  \bibinfo{pages}{094509} (\bibinfo{year}{2012}).

\bibitem[{\citenamefont{Allan et~al.}(2012)\citenamefont{Allan, Rost,
  Mackenzie, Xie, Davis, Lee, Iyo., Eisaki, and Chuang}}]{all2012}
\bibinfo{author}{\bibfnamefont{M.~P.} \bibnamefont{Allan}},
  \bibinfo{author}{\bibfnamefont{A.~W.} \bibnamefont{Rost}},
  \bibinfo{author}{\bibfnamefont{A.~P.} \bibnamefont{Mackenzie}},
  \bibinfo{author}{\bibfnamefont{Y.}~\bibnamefont{Xie}},
  \bibinfo{author}{\bibfnamefont{J.~C.} \bibnamefont{Davis}},
  \bibinfo{author}{\bibfnamefont{K.~K. C.~H.} \bibnamefont{Lee}},
  \bibinfo{author}{\bibfnamefont{A.}~\bibnamefont{Iyo.}},
  \bibinfo{author}{\bibfnamefont{H.}~\bibnamefont{Eisaki}}, \bibnamefont{and}
  \bibinfo{author}{\bibfnamefont{T.-M.} \bibnamefont{Chuang}},
  \bibinfo{journal}{Science} \textbf{\bibinfo{volume}{336}},
  \bibinfo{pages}{563} (\bibinfo{year}{2012}).

\bibitem[{\citenamefont{Jang et~al.}(2012)\citenamefont{Jang, Hong, Kwon, Park,
  Gofryk, Ronning, Thompson, and Bang}}]{jan2012}
\bibinfo{author}{\bibfnamefont{D.-J.} \bibnamefont{Jang}},
  \bibinfo{author}{\bibfnamefont{J.~B.} \bibnamefont{Hong}},
  \bibinfo{author}{\bibfnamefont{Y.~S.} \bibnamefont{Kwon}},
  \bibinfo{author}{\bibfnamefont{T.}~\bibnamefont{Park}},
  \bibinfo{author}{\bibfnamefont{K.}~\bibnamefont{Gofryk}},
  \bibinfo{author}{\bibfnamefont{F.}~\bibnamefont{Ronning}},
  \bibinfo{author}{\bibfnamefont{J.~D.} \bibnamefont{Thompson}},
  \bibnamefont{and} \bibinfo{author}{\bibfnamefont{Y.}~\bibnamefont{Bang}},
  \bibinfo{journal}{Phys. Rev. B} \textbf{\bibinfo{volume}{85}},
  \bibinfo{pages}{180505(R)} (\bibinfo{year}{2012}).

\bibitem[{\citenamefont{Platt et~al.}(2011)\citenamefont{Platt, Thomale, and
  Hanke}}]{pla2011}
\bibinfo{author}{\bibfnamefont{C.}~\bibnamefont{Platt}},
  \bibinfo{author}{\bibfnamefont{R.}~\bibnamefont{Thomale}}, \bibnamefont{and}
  \bibinfo{author}{\bibfnamefont{W.}~\bibnamefont{Hanke}},
  \bibinfo{journal}{Phys. Rev. B} \textbf{\bibinfo{volume}{84}},
  \bibinfo{pages}{235121} (\bibinfo{year}{2011}).

\bibitem[{\citenamefont{H\"anke et~al.}(2012)\citenamefont{H\"anke, Sykora,
  Schlegel, Baumann, Harnagea, Wurmehl, Daghofer, B\"uchner, van~den Brink, and
  Hess}}]{han2012}
\bibinfo{author}{\bibfnamefont{T.}~\bibnamefont{H\"anke}},
  \bibinfo{author}{\bibfnamefont{S.}~\bibnamefont{Sykora}},
  \bibinfo{author}{\bibfnamefont{R.}~\bibnamefont{Schlegel}},
  \bibinfo{author}{\bibfnamefont{D.}~\bibnamefont{Baumann}},
  \bibinfo{author}{\bibfnamefont{L.}~\bibnamefont{Harnagea}},
  \bibinfo{author}{\bibfnamefont{S.}~\bibnamefont{Wurmehl}},
  \bibinfo{author}{\bibfnamefont{M.}~\bibnamefont{Daghofer}},
  \bibinfo{author}{\bibfnamefont{B.}~\bibnamefont{B\"uchner}},
  \bibinfo{author}{\bibfnamefont{J.}~\bibnamefont{van~den Brink}},
  \bibnamefont{and} \bibinfo{author}{\bibfnamefont{C.}~\bibnamefont{Hess}},
  \bibinfo{journal}{Phys. Rev. Lett} \textbf{\bibinfo{volume}{108}},
  \bibinfo{pages}{127001} (\bibinfo{year}{2012}).

\bibitem[{\citenamefont{Baek et~al.}(2012)\citenamefont{Baek, Grafe.,
  Hammerath, Fuchs, Rudisch, Harnagea, Aswartham, Wurmehl, van~den Brink, and
  B\"uchner}}]{bae2012}
\bibinfo{author}{\bibfnamefont{S.~H.} \bibnamefont{Baek}},
  \bibinfo{author}{\bibfnamefont{H.~J.} \bibnamefont{Grafe.}},
  \bibinfo{author}{\bibfnamefont{F.}~\bibnamefont{Hammerath}},
  \bibinfo{author}{\bibfnamefont{M.}~\bibnamefont{Fuchs}},
  \bibinfo{author}{\bibfnamefont{C.}~\bibnamefont{Rudisch}},
  \bibinfo{author}{\bibfnamefont{L.}~\bibnamefont{Harnagea}},
  \bibinfo{author}{\bibfnamefont{S.}~\bibnamefont{Aswartham}},
  \bibinfo{author}{\bibfnamefont{S.}~\bibnamefont{Wurmehl}},
  \bibinfo{author}{\bibfnamefont{J.}~\bibnamefont{van~den Brink}},
  \bibnamefont{and}
  \bibinfo{author}{\bibfnamefont{B.}~\bibnamefont{B\"uchner}},
  \bibinfo{journal}{Eur. Phys. J. B} \textbf{\bibinfo{volume}{85}},
  \bibinfo{pages}{159} (\bibinfo{year}{2012}).

\bibitem[{\citenamefont{Baek et~al.}(2013)\citenamefont{Baek, Harnagea,
  Wurmehl., B\"uchner, and Grafe}}]{bae2013}
\bibinfo{author}{\bibfnamefont{S.~H.} \bibnamefont{Baek}},
  \bibinfo{author}{\bibfnamefont{L.}~\bibnamefont{Harnagea}},
  \bibinfo{author}{\bibfnamefont{S.}~\bibnamefont{Wurmehl.}},
  \bibinfo{author}{\bibfnamefont{B.}~\bibnamefont{B\"uchner}},
  \bibnamefont{and} \bibinfo{author}{\bibfnamefont{H.~J.} \bibnamefont{Grafe}},
  \bibinfo{journal}{J. Phys.: Condens. Matter} \textbf{\bibinfo{volume}{25}},
  \bibinfo{pages}{162204} (\bibinfo{year}{2013}).

\bibitem[{\citenamefont{Qureshi
  et~al.}(2012{\natexlab{a}})\citenamefont{Qureshi, Steffens, Drees, Komarek,
  Lamago, Sidis, Harnagea, Grafe, Wurmehl, B\"uchner et~al.}}]{qur2012}
\bibinfo{author}{\bibfnamefont{N.}~\bibnamefont{Qureshi}},
  \bibinfo{author}{\bibfnamefont{P.}~\bibnamefont{Steffens}},
  \bibinfo{author}{\bibfnamefont{Y.}~\bibnamefont{Drees}},
  \bibinfo{author}{\bibfnamefont{A.~C.} \bibnamefont{Komarek}},
  \bibinfo{author}{\bibfnamefont{D.}~\bibnamefont{Lamago}},
  \bibinfo{author}{\bibfnamefont{Y.}~\bibnamefont{Sidis}},
  \bibinfo{author}{\bibfnamefont{L.}~\bibnamefont{Harnagea}},
  \bibinfo{author}{\bibfnamefont{H.-J.} \bibnamefont{Grafe}},
  \bibinfo{author}{\bibfnamefont{S.}~\bibnamefont{Wurmehl}},
  \bibinfo{author}{\bibfnamefont{B.}~\bibnamefont{B\"uchner}},
  \bibnamefont{et~al.}, \bibinfo{journal}{Phys. Rev. Lett.}
  \textbf{\bibinfo{volume}{108}}, \bibinfo{pages}{117001}
  (\bibinfo{year}{2012}{\natexlab{a}}).

\bibitem[{\citenamefont{Wang et~al.}(2012)\citenamefont{Wang, Wang, Miao, Carr,
  Abernathy, Stone, Wang, Xing, Jin, Zhang et~al.}}]{wan2012}
\bibinfo{author}{\bibfnamefont{M.}~\bibnamefont{Wang}},
  \bibinfo{author}{\bibfnamefont{M.}~\bibnamefont{Wang}},
  \bibinfo{author}{\bibfnamefont{H.}~\bibnamefont{Miao}},
  \bibinfo{author}{\bibfnamefont{S.}~\bibnamefont{Carr}},
  \bibinfo{author}{\bibfnamefont{D.~L.} \bibnamefont{Abernathy}},
  \bibinfo{author}{\bibfnamefont{M.~B.} \bibnamefont{Stone}},
  \bibinfo{author}{\bibfnamefont{X.~C.} \bibnamefont{Wang}},
  \bibinfo{author}{\bibfnamefont{L.}~\bibnamefont{Xing}},
  \bibinfo{author}{\bibfnamefont{C.~Q.} \bibnamefont{Jin}},
  \bibinfo{author}{\bibfnamefont{X.}~\bibnamefont{Zhang}},
  \bibnamefont{et~al.}, \bibinfo{journal}{Phys. Rev. B}
  \textbf{\bibinfo{volume}{86}}, \bibinfo{pages}{144511}
  (\bibinfo{year}{2012}).

\bibitem[{\citenamefont{Knolle et~al.}(2012)\citenamefont{Knolle, Zabolotnyy,
  Eremin, Borisenko, Qureshi, Braden, Evtushinsky, Kim, Kordyuk, Sykora
  et~al.}}]{kno2012}
\bibinfo{author}{\bibfnamefont{J.}~\bibnamefont{Knolle}},
  \bibinfo{author}{\bibfnamefont{V.~B.} \bibnamefont{Zabolotnyy}},
  \bibinfo{author}{\bibfnamefont{I.}~\bibnamefont{Eremin}},
  \bibinfo{author}{\bibfnamefont{S.~V.} \bibnamefont{Borisenko}},
  \bibinfo{author}{\bibfnamefont{N.}~\bibnamefont{Qureshi}},
  \bibinfo{author}{\bibfnamefont{M.}~\bibnamefont{Braden}},
  \bibinfo{author}{\bibfnamefont{D.~V.} \bibnamefont{Evtushinsky}},
  \bibinfo{author}{\bibfnamefont{T.~K.} \bibnamefont{Kim}},
  \bibinfo{author}{\bibfnamefont{A.~A.} \bibnamefont{Kordyuk}},
  \bibinfo{author}{\bibfnamefont{S.}~\bibnamefont{Sykora}},
  \bibnamefont{et~al.}, \bibinfo{journal}{Phys. Rev. B}
  \textbf{\bibinfo{volume}{86}}, \bibinfo{pages}{174519}
  (\bibinfo{year}{2012}).

\bibitem[{\citenamefont{Xu et~al.}(2013)\citenamefont{Xu, Xu, and
  Tranquada}}]{xu2013}
\bibinfo{author}{\bibfnamefont{G.}~\bibnamefont{Xu}},
  \bibinfo{author}{\bibfnamefont{Z.}~\bibnamefont{Xu}}, \bibnamefont{and}
  \bibinfo{author}{\bibfnamefont{J.~M.} \bibnamefont{Tranquada}},
  \bibinfo{journal}{Rev. Sci. Instrum.} \textbf{\bibinfo{volume}{84}},
  \bibinfo{pages}{083906} (\bibinfo{year}{2013}).

\bibitem[{\citenamefont{Zheludev}(2006)}]{reslib}
\bibinfo{author}{\bibfnamefont{A.}~\bibnamefont{Zheludev}},
  \bibinfo{journal}{{\sc reslib} 3.4c (Oak Ridge National Laboratory, Oak
  Ridge, TN)}  (\bibinfo{year}{2006}).

\bibitem[{\citenamefont{Popovici}(1975)}]{pop1975}
\bibinfo{author}{\bibfnamefont{M.}~\bibnamefont{Popovici}},
  \bibinfo{journal}{Acta Crystallogr., Sect. A: Found. Crystallogr.}
  \textbf{\bibinfo{volume}{31}}, \bibinfo{pages}{507} (\bibinfo{year}{1975}).

\bibitem[{\citenamefont{Marshall and Lovesey}(1971)}]{mar1971}
\bibinfo{author}{\bibfnamefont{W.}~\bibnamefont{Marshall}} \bibnamefont{and}
  \bibinfo{author}{\bibfnamefont{S.}~\bibnamefont{Lovesey}},
  \emph{\bibinfo{title}{Theory of thermal Neutron Scattering}}
  (\bibinfo{publisher}{Oxford University Press}, \bibinfo{year}{1971}).

\bibitem[{\citenamefont{Squires}(1997)}]{squ1997}
\bibinfo{author}{\bibfnamefont{G.~L.} \bibnamefont{Squires}},
  \emph{\bibinfo{title}{Introduction to the theory of thermal neutron
  scattering}} (\bibinfo{publisher}{Dover Publications Inc.},
  \bibinfo{year}{1997}).

\bibitem[{\citenamefont{Yin et~al.}(2011)\citenamefont{Yin, Haule, and
  Kotliar}}]{yin2011}
\bibinfo{author}{\bibfnamefont{Z.~P.} \bibnamefont{Yin}},
  \bibinfo{author}{\bibfnamefont{K.}~\bibnamefont{Haule}}, \bibnamefont{and}
  \bibinfo{author}{\bibfnamefont{G.}~\bibnamefont{Kotliar}},
  \bibinfo{journal}{Nat. Mater.} \textbf{\bibinfo{volume}{10}},
  \bibinfo{pages}{932} (\bibinfo{year}{2011}).

\bibitem[{\citenamefont{Ferber et~al.}(2012)\citenamefont{Ferber, Foyevtsova,
  Valent\'i, and Jeschke}}]{fer2012}
\bibinfo{author}{\bibfnamefont{J.}~\bibnamefont{Ferber}},
  \bibinfo{author}{\bibfnamefont{K.}~\bibnamefont{Foyevtsova}},
  \bibinfo{author}{\bibfnamefont{R.}~\bibnamefont{Valent\'i}},
  \bibnamefont{and} \bibinfo{author}{\bibfnamefont{H.~O.}
  \bibnamefont{Jeschke}}, \bibinfo{journal}{Phys. Rev. B}
  \textbf{\bibinfo{volume}{85}}, \bibinfo{pages}{094505}
  (\bibinfo{year}{2012}).

\bibitem[{\citenamefont{Braden et~al.}(2002)\citenamefont{Braden, Sidis,
  Bourges, Pfeuty, Kulda, Mao, and Maeno}}]{bra2002}
\bibinfo{author}{\bibfnamefont{M.}~\bibnamefont{Braden}},
  \bibinfo{author}{\bibfnamefont{Y.}~\bibnamefont{Sidis}},
  \bibinfo{author}{\bibfnamefont{P.}~\bibnamefont{Bourges}},
  \bibinfo{author}{\bibfnamefont{P.}~\bibnamefont{Pfeuty}},
  \bibinfo{author}{\bibfnamefont{J.}~\bibnamefont{Kulda}},
  \bibinfo{author}{\bibfnamefont{Z.}~\bibnamefont{Mao}}, \bibnamefont{and}
  \bibinfo{author}{\bibfnamefont{Y.}~\bibnamefont{Maeno}},
  \bibinfo{journal}{Phys. Rev. B} \textbf{\bibinfo{volume}{66}},
  \bibinfo{pages}{064522} (\bibinfo{year}{2002}).

\bibitem[{\citenamefont{Wang et~al.}(2013)\citenamefont{Wang, Kreisel,
  Zabolotnyy, Borisenko, B\"uchner, Maier, Hirschfeld, and
  Scalapino}}]{wan2013}
\bibinfo{author}{\bibfnamefont{Y.}~\bibnamefont{Wang}},
  \bibinfo{author}{\bibfnamefont{A.}~\bibnamefont{Kreisel}},
  \bibinfo{author}{\bibfnamefont{V.~B.} \bibnamefont{Zabolotnyy}},
  \bibinfo{author}{\bibfnamefont{S.~V.} \bibnamefont{Borisenko}},
  \bibinfo{author}{\bibfnamefont{B.}~\bibnamefont{B\"uchner}},
  \bibinfo{author}{\bibfnamefont{T.~A.} \bibnamefont{Maier}},
  \bibinfo{author}{\bibfnamefont{P.~J.} \bibnamefont{Hirschfeld}},
  \bibnamefont{and} \bibinfo{author}{\bibfnamefont{D.~J.}
  \bibnamefont{Scalapino}}, \bibinfo{journal}{Phys. Rev. B}
  \textbf{\bibinfo{volume}{88}}, \bibinfo{pages}{174516}
  (\bibinfo{year}{2013}).

\bibitem[{\citenamefont{Yin et~al.}()\citenamefont{Yin, Haule, and
  Kotliar}}]{yin2013}
\bibinfo{author}{\bibfnamefont{Z.~P.} \bibnamefont{Yin}},
  \bibinfo{author}{\bibfnamefont{K.}~\bibnamefont{Haule}}, \bibnamefont{and}
  \bibinfo{author}{\bibfnamefont{G.}~\bibnamefont{Kotliar}},
  \bibinfo{note}{arXiv:1311.1188}.

\bibitem[{\citenamefont{Ahn et~al.}()\citenamefont{Ahn, Eremin, Knolle,
  Zabolotnyy, Borisenko, B\"uchner, and Chubukov}}]{ahn2014}
\bibinfo{author}{\bibfnamefont{F.}~\bibnamefont{Ahn}},
  \bibinfo{author}{\bibfnamefont{I.}~\bibnamefont{Eremin}},
  \bibinfo{author}{\bibfnamefont{J.}~\bibnamefont{Knolle}},
  \bibinfo{author}{\bibfnamefont{V.~B.} \bibnamefont{Zabolotnyy}},
  \bibinfo{author}{\bibfnamefont{S.~V.} \bibnamefont{Borisenko}},
  \bibinfo{author}{\bibfnamefont{B.}~\bibnamefont{B\"uchner}},
  \bibnamefont{and} \bibinfo{author}{\bibfnamefont{A.~V.}
  \bibnamefont{Chubukov}}, \bibinfo{note}{arXiv:1402.2112}.

\bibitem[{\citenamefont{Saito et~al.}()\citenamefont{Saito, Onari, Yamakawa,
  Kontani, Borisenko, and Zabolotnyy}}]{sai2014}
\bibinfo{author}{\bibfnamefont{T.}~\bibnamefont{Saito}},
  \bibinfo{author}{\bibfnamefont{S.}~\bibnamefont{Onari}},
  \bibinfo{author}{\bibfnamefont{Y.}~\bibnamefont{Yamakawa}},
  \bibinfo{author}{\bibfnamefont{H.}~\bibnamefont{Kontani}},
  \bibinfo{author}{\bibfnamefont{S.~V.} \bibnamefont{Borisenko}},
  \bibnamefont{and} \bibinfo{author}{\bibfnamefont{V.~B.}
  \bibnamefont{Zabolotnyy}}, \bibinfo{note}{arXiv:1402.2398}.

\bibitem[{\citenamefont{Qureshi
  et~al.}(2012{\natexlab{b}})\citenamefont{Qureshi, Steffens, Wurmehl,
  Aswartham, B\"uchner, and Braden}}]{qur2012b}
\bibinfo{author}{\bibfnamefont{N.}~\bibnamefont{Qureshi}},
  \bibinfo{author}{\bibfnamefont{P.}~\bibnamefont{Steffens}},
  \bibinfo{author}{\bibfnamefont{S.}~\bibnamefont{Wurmehl}},
  \bibinfo{author}{\bibfnamefont{S.}~\bibnamefont{Aswartham}},
  \bibinfo{author}{\bibfnamefont{B.}~\bibnamefont{B\"uchner}},
  \bibnamefont{and} \bibinfo{author}{\bibfnamefont{M.}~\bibnamefont{Braden}},
  \bibinfo{journal}{Phys. Rev. B} \textbf{\bibinfo{volume}{86}},
  \bibinfo{pages}{060410(R)} (\bibinfo{year}{2012}{\natexlab{b}}).

\bibitem[{\citenamefont{Steffens et~al.}(2013)\citenamefont{Steffens, Lee,
  Qureshi, Kihou, Iyo, Eisaki, and Braden}}]{ste2013}
\bibinfo{author}{\bibfnamefont{P.}~\bibnamefont{Steffens}},
  \bibinfo{author}{\bibfnamefont{C.~H.} \bibnamefont{Lee}},
  \bibinfo{author}{\bibfnamefont{N.}~\bibnamefont{Qureshi}},
  \bibinfo{author}{\bibfnamefont{K.}~\bibnamefont{Kihou}},
  \bibinfo{author}{\bibfnamefont{A.}~\bibnamefont{Iyo}},
  \bibinfo{author}{\bibfnamefont{H.}~\bibnamefont{Eisaki}}, \bibnamefont{and}
  \bibinfo{author}{\bibfnamefont{M.}~\bibnamefont{Braden}},
  \bibinfo{journal}{Phys. Rev. Lett.} \textbf{\bibinfo{volume}{110}},
  \bibinfo{pages}{137001} (\bibinfo{year}{2013}).

\bibitem[{\citenamefont{Liu et~al.}(2012)\citenamefont{Liu, Lester, Kulda, Lu,
  Luo, Wang, Hayden, and Dai}}]{liu2012}
\bibinfo{author}{\bibfnamefont{M.}~\bibnamefont{Liu}},
  \bibinfo{author}{\bibfnamefont{C.}~\bibnamefont{Lester}},
  \bibinfo{author}{\bibfnamefont{J.}~\bibnamefont{Kulda}},
  \bibinfo{author}{\bibfnamefont{X.}~\bibnamefont{Lu}},
  \bibinfo{author}{\bibfnamefont{H.}~\bibnamefont{Luo}},
  \bibinfo{author}{\bibfnamefont{M.}~\bibnamefont{Wang}},
  \bibinfo{author}{\bibfnamefont{S.~M.} \bibnamefont{Hayden}},
  \bibnamefont{and} \bibinfo{author}{\bibfnamefont{P.}~\bibnamefont{Dai}},
  \bibinfo{journal}{Phys. Rev. B} \textbf{\bibinfo{volume}{85}},
  \bibinfo{pages}{214516} (\bibinfo{year}{2012}).

\bibitem[{\citenamefont{Luo et~al.}(2013)\citenamefont{Luo, Wang, Zhang, Lu,
  Regnault, Zhang, Li, Hu, and Dai}}]{luo2013}
\bibinfo{author}{\bibfnamefont{H.}~\bibnamefont{Luo}},
  \bibinfo{author}{\bibfnamefont{M.}~\bibnamefont{Wang}},
  \bibinfo{author}{\bibfnamefont{C.}~\bibnamefont{Zhang}},
  \bibinfo{author}{\bibfnamefont{X.}~\bibnamefont{Lu}},
  \bibinfo{author}{\bibfnamefont{L.-P.} \bibnamefont{Regnault}},
  \bibinfo{author}{\bibfnamefont{R.}~\bibnamefont{Zhang}},
  \bibinfo{author}{\bibfnamefont{S.}~\bibnamefont{Li}},
  \bibinfo{author}{\bibfnamefont{J.}~\bibnamefont{Hu}}, \bibnamefont{and}
  \bibinfo{author}{\bibfnamefont{P.}~\bibnamefont{Dai}},
  \bibinfo{journal}{Phys. Rev. Lett.} \textbf{\bibinfo{volume}{111}},
  \bibinfo{pages}{107006} (\bibinfo{year}{2013}).

\bibitem[{\citenamefont{Taylor et~al.}(2011)\citenamefont{Taylor, Pitcher,
  Ewings, Perring, Clarke, and Boothroyd}}]{tay2011}
\bibinfo{author}{\bibfnamefont{A.~E.} \bibnamefont{Taylor}},
  \bibinfo{author}{\bibfnamefont{M.~J.} \bibnamefont{Pitcher}},
  \bibinfo{author}{\bibfnamefont{R.~A.} \bibnamefont{Ewings}},
  \bibinfo{author}{\bibfnamefont{T.~G.} \bibnamefont{Perring}},
  \bibinfo{author}{\bibfnamefont{S.~J.} \bibnamefont{Clarke}},
  \bibnamefont{and} \bibinfo{author}{\bibfnamefont{A.~T.}
  \bibnamefont{Boothroyd}}, \bibinfo{journal}{Phys. Rev. B}
  \textbf{\bibinfo{volume}{83}}, \bibinfo{pages}{220514(R)}
  (\bibinfo{year}{2011}).

\bibitem[{\citenamefont{Lester et~al.}(2010)\citenamefont{Lester, Chu,
  Analytis, Perring, Fisher, and Hayden}}]{les2010}
\bibinfo{author}{\bibfnamefont{C.}~\bibnamefont{Lester}},
  \bibinfo{author}{\bibfnamefont{J.-H.} \bibnamefont{Chu}},
  \bibinfo{author}{\bibfnamefont{J.~G.} \bibnamefont{Analytis}},
  \bibinfo{author}{\bibfnamefont{T.~G.} \bibnamefont{Perring}},
  \bibinfo{author}{\bibfnamefont{I.~R.} \bibnamefont{Fisher}},
  \bibnamefont{and} \bibinfo{author}{\bibfnamefont{S.~M.}
  \bibnamefont{Hayden}}, \bibinfo{journal}{Phys. Rev. B}
  \textbf{\bibinfo{volume}{81}}, \bibinfo{pages}{064505}
  (\bibinfo{year}{2010}).

\end{thebibliography}
\end{document}